\documentclass[12pt]{article}
\newcommand{\apj}{Astrophys. J}
\newcommand{\apjs}{Astrophys. J. Suppl.}
\newcommand{\prd}{Phys. Rev. D}

\begin{document}
\def\stackunder#1#2{\mathrel{\mathop{#2}\limits_{#1}}}

\title{Covariant Linear Perturbations in a Concordance Model}

\author{\small Viktor Czinner, M\'aty\'as Vas\'uth, \'Arp\'ad Luk\'acs and \framebox{Zolt\'an Perj\'es}\\
\small KFKI Research Institute for Particle and Nuclear Physics,\\
\small Budapest 114, P.O.Box 49, H-1525 Hungary\\ 
\small czinner@rmki.kfki.hu, vasuth@rmki.kfki.hu, arpi@rmki.kfki.hu}
\date{}
\maketitle

\begin{abstract}
We present the complete solution of the first order metric and density perturbation equations in a
spatially flat ($K=0$), Friedmann-Robertson-Walker (FRW) universe filled with pressureless ideal 
fluid, in the presence of cosmological constant. We use covariant linear perturbation formalism 
and the comoving gauge condition to obtain the field and conservation equations. The solution 
contains all modes of the perturbations, i.e. scalar, vector and tensor modes, and we show that 
our results are in agreement with the Sachs \& Wolfe metric perturbation formalism.     
\end{abstract}
\vskip 12pt


\section{Introduction}
Recent cosmological measurements by the {\it Wilkinson Microvawe Anisotropy Probe} (WMAP) \cite{Spergel}, 
Ia type supernovae observations \cite{Tonry} and other sky surveys eg. {\it Sloan Digital Sky Survey} (SDSS) 
\cite{Tegmark} seem to prove today that the expansion of the universe is accelerating. Although there 
exist many different approaches explaining this phenomenon, the most accepted description in the literature 
is the {\it concordance} model. In this model the universe is flat $(K=0)$, homogeneous 
and isotropic with nonzero cosmological constant $\Lambda$ in the Einstein equations. 

In this work our aim is to obtain the complete first order solution of the perturbations of a concordance
model filled with pressureless matter in Bardeen's covariant formalism \cite{Bardeen}. 
Hu \& Sugiyama \cite{HS} claim that in the presence of cosmological constant one must use a numerical 
approach to get the result. Recent calculations of Perj\'es et al. \cite{PVCE} yield all the $C^{\infty}$ 
solution of the problem following the original work of Sachs \& Wolfe \cite{SW}. Here we show that 
the complete analytic solution of the perturbed field equations can also be obtained using 
the covariant linear perturbation formalism, in agreement with the results of Perj\'es et al. \cite{PVCE}. 
Providing the solution of the problem in this formalism has the advantage that recent model calculations of 
the fluctuations of the Cosmic Microwave Background Radiation (CMBR) prefer using Bardeen's covariant linear 
formalism rather than the perturbation approach of Sachs \& Wolfe. 

In Sec. 2, we present the well known solution of the background quantities in a flat, homogeneous and 
isotropic universe and give a brief description of the covariant linear perturbation formalism. 
The perturbed field equations and their solution for scalar, vector and tensor modes are presented in 
Sec. 3. In Sec. 4 we build up the complete first order solution of the metric and show the agreement 
with the results of \cite{PVCE}.

\section{Background metric and linear perturbations in covariant formalism}
As background solution we use the homogeneous and isotropic FRW metric 
\begin{eqnarray} \label{ds2}
ds^2 = a^{2}(\eta)(-d\eta^2+\gamma_{\mu\nu}dx^\mu dx^\nu) \ ,
\end{eqnarray}
where $\gamma_{\mu\nu}$ is the three metric of a space with constant spatial curvature $K$, 
and $\eta$ is the conformal time variable. Introducing the comoving time coordinate $t$ by 
the relation $ad\eta=dt$, the scale factor $a(t)$ can be expressed in the case of a flat 
($K=0$) space and pressureless ($p=0$) matter source as~\cite{Stephani}
\begin{eqnarray}  \label{backgr}
a=a_0 {\rm sinh}^{2/3}(Ct+C_0)\ ,\quad  \rho a^3={\cal C}_M \ , \quad 
a_0=\left(\frac{{\cal C}_M}{\Lambda}\right)^{1/3}\ ,\quad 
C=\frac{\sqrt{3\Lambda}}{2} \ .
\end{eqnarray}
Throughout this paper we use units in which the gravitational constant $G=1/8\pi$, the speed of light $c=1$
and we set $C_0=0$. Roman indices run from 0 to 3 and Greek indices run from 1 to 3. We use $\gamma_{\mu\nu}$
and its inverse to raise and lower spatial indices of first order quantities.

To obtain the perturbed field equations we use Bardeen's covariant linear formalism~\cite{Bardeen} as it is 
presented by Hu \cite{Hu}. Hereafter we discuss only the main steps of this description to derive the basic 
equations. 

For a spatially flat FRW universe the general perturbations of the metric tensor have the following form 
\begin{eqnarray} \label{metric}
g_{00} &=& -a^{2}(1+2A ) \ , \nonumber\\
g_{0\alpha} &=& -a^{2} B_\alpha \ , \\
g_{\alpha\beta} &=& a^{2} \left[\left(1+2H_L\right)\gamma_{\alpha\beta} 
 + 2H_{T\alpha\beta}\right] \ .\nonumber
\end{eqnarray}
The functions $A$, $H_L$, $B_\alpha$ and $H_{T\alpha\beta}$ give a complete representation
of the metric, where $H_{T\alpha\beta}$ is a $3\times 3$ trace-free tensor.  

We define an effective energy momentum tensor in the Einstein equations with cosmological constant as  
\begin{eqnarray}
R_{ik}-\frac{1}{2}Rg_{ik}=\tilde T_{ik}=T_{ik}+\Lambda g_{ik}\ . 
\end{eqnarray}
The energy momentum tensor $\tilde T_{ik}$ in the case of a homogeneous and isotropic ideal fluid source can be 
generally perturbed as
\begin{eqnarray} \label{energy-moment}
\tilde{T}^0_{\ 0} &=& -\tilde\rho-\delta\rho \ , \nonumber\\
\tilde{T}^0_{\ \alpha} &=& (\tilde\rho+\tilde p)(v_\alpha-B_\alpha) \ , \\
\tilde{T}^\alpha_{\ 0} &=& -(\tilde\rho+\tilde p)v^\alpha \ , \nonumber \\
\tilde{T}^\alpha_{\ \beta} &=& (\tilde p+\delta p)\delta^\alpha_{\ \beta}+\tilde p\Pi^\alpha_{\ \beta} \ ,\nonumber
\end{eqnarray}
where $\delta\rho$, $\delta p$ and $v_\alpha$ are the density, pressure and velocity perturbations, respectively,
and we introduce the effective energy density $\tilde\rho$ and pressure $\tilde p$ as new variables with the following 
definition
\begin{equation}
\tilde\rho=\rho+\Lambda\ ,\quad \tilde p=p-\Lambda\ .
\end{equation}
The $\Pi^\alpha_{\ \beta}$ tensor represents the anisotropic stress perturbations. 

In covariant linear formalism the Einstein equations can be decoupled into a set of ordinary differential 
equations by employing scalar, vector and tensor eigenmodes of the Laplacian operator which form a complete 
set. In a spatially flat ($K=0$) universe, these eigenmodes are plane waves 
\begin{eqnarray}\label{eigenmodes}
{ Q^{(0)} } &=& { \exp( i {\bf k} \cdot {\bf x}) } \,,  \nonumber\\
{ Q_{\alpha}^{(\pm 1)} }
	&=& \frac{-i}{\sqrt{2}} (\hat{\bf e}_1 \pm i \hat{\bf e}_2)_{\alpha} 
	{ \exp(i {\bf k}\cdot {\bf x})\,, }\\
{ Q_{\alpha\beta}^{(\pm 2)} }
 	&=& - \sqrt{\frac{3}{8}}
(\hat{\bf e}_1 \pm i \hat{\bf e}_2)_{\alpha} (\hat{\bf e}_1 \pm i \hat{\bf e}_2)_{\beta}
{\exp(i {\bf k}\cdot {\bf x})}\,,\nonumber 
 \end{eqnarray}
with the unit vectors ${\bf e}_1$ and ${\bf e}_2$ spanning the plane transverse to the wave vector ${\bf k}$.
For an arbitrary scalar, vector and tensor function the components of the $k$th eigenmode become
\begin{equation} 
F({\bf x}) = F (k)\, Q^{(0)},\ F_{\alpha}({\bf x}) = \!\!\!\sum_{m=-1}^1 F^{(m)}(k) \, Q_{\alpha}^{(m)},\
F_{\alpha\beta}({\bf x})  = \!\!\! \sum_{m=-2}^2  F^{(m)}(k) \,  Q_{\alpha\beta}^{(m)}.
\end{equation}
In the plane wave expansion there are relations between the curl free vectors and longitudinal components of tensors, 
and the scalar and vector modes of the eigenfunctions as follows 
\begin{eqnarray}
Q_{\alpha}^{(0)} & = & -k^{-1} \nabla_{\alpha} Q^{(0)}, \nonumber \\
Q_{\alpha\beta}^{(0)}&=&(k^{-2} \nabla_{\alpha} \nabla_{\beta} + {1 \over 3} \gamma_{\alpha\beta}) Q^{(0)},  \\
Q_{\alpha\beta}^{(\pm 1)}&=&-{1 \over 2k}[ \nabla_{\alpha} Q_{\beta}^{(\pm 1)}+ \nabla_{\beta} Q_{\alpha}^{(\pm 1)}],\nonumber  
\end{eqnarray}
where $\nabla$ is the covariant derivative operator with respect to $\gamma_{\alpha\beta}$.

In the perturbed space-time we choose comoving coordinates. Using this gauge the following conditions hold  
\begin{eqnarray} \label{comov}
A=0 \ , \quad v^\alpha=0 \ .
\end{eqnarray}
In further calculations we are interested in a spatially flat ($K=0$) universe filled with pressureless ($p=0$) 
ideal fluid and vanishing anisotropic stress perturbations ($\Pi^\alpha_{\ \beta}=0$).

\section{The field equations and their solution}
In this section we solve the Einstein equations for scalar, vector and tensor modes. First we present
the field equations using the conditions discussed at the end of the previous section, then we 
give the complete solution for each mode of the perturbations. \\\\
{\it Scalar modes}\\\\
The field equations for scalar modes have the following form 
\begin{eqnarray}\label{seinst} 
 k^2\left[  {H_L}+ {1 \over 3}  {H_T}+ \dot a \left({ {B} \over k} \right.\right.&-&\left.\left.{ a\dot H_T \over k^{2} } \right)\right]  
	= \frac{a^2}{2}  \left[{\delta \rho} - 3 \dot a  \rho{{B} \over k}\right] \ , \nonumber\\
k^2 \left({H_L} + {1 \over 3}  {H_T} \right) &+& \left(a {d \over dt}+ 2 \dot a  \right)
(k  {B} -   {a\dot H_T}) = 0  \ ,\\
{a\dot H_L} +{1 \over 3}  {a\dot H_T} =  \frac{a^2}{2} \rho{{B} \over k} \ &,&\quad 
 \left[ a { d\over dt} + \dot a \right] \left( {a \dot H_L} +  {k  {B} \over 3}\right)
=- \frac{a^2}{6} \delta\rho \ ,\nonumber
\end{eqnarray}
and the scalar modes of the conservation equations become 
\begin{eqnarray} \label{sdiv}
\left(a{d \over dt} + 3 \dot a \right)  {\delta\rho}
	 =-3\rho  {a\dot H_L}\ ,\quad \left( a{d \over dt} + 4\dot a \right) \rho{{B} \over k} = 0\ .
\end{eqnarray}
For the scalar quantities the solution of the field equations gives
\begin{eqnarray} \label{seqs}
\delta\rho &=& \frac{\cosh(Ct)}{\sinh^3(Ct)}\left[K_1(k)-K_2(k) I(t)\right] \ ,\quad
B=\frac{k}{C_M}\frac{B_0(k)}{a} \ , \nonumber\\
H_L &=& -\frac{a_0^3}{3C_M}\coth(Ct)\left[K_1(k)-K_2(k) I(t)\right]+V(k) \ , \\
H_T &=& -3H_L-\frac{3B_0(k)}{2Ca_0^3}\coth(Ct)+H_0(k) \ , \nonumber
\end{eqnarray}
where $K_1(k)$, $K_2(k)$, $B_0(k)$, $V(k)$ and $H_0(k)$ are arbitrary scalar functions
depending on spatial coordinates. The function $I(t)$ is the following integral of the time variable
\begin{eqnarray}
&&I(t)=2^{-2/3}\sqrt{3\Lambda}\int^{\ \!t}_0 \frac{{\rm sinh}^{2/3}(C\tau)}{{\rm cosh}^2(C\tau)}\ {\rm d}\tau \ .
\end{eqnarray}  
The Legendre normal form of the elliptic integral $I(t)$ is given in the appendix of \cite{PVCE}.\newpage 
\noindent{\it Vector modes}\\\\
The Einstein equations for vector modes are 
\begin{eqnarray} \label{veinst}
\left(k {B^{(\pm 1)}} -  {a {\dot H}_T}^{(\pm 1)}\right)&=&-2 a^2 \rho {{B^{(\pm 1)}} \over k} \ ,\\
\left[ a{d \over dt} + 2 \dot a \right] \left(k {B^{(\pm 1)}} \right.
&-&\left. {a{\dot H}_T}^{(\pm 1)}\right) = 0\label{veinst1}
\end{eqnarray}\
and the conservation equations become
\begin{eqnarray} \label{vdiv}
\left[a{d \over dt}+4 \dot a\right]\rho{{B}^{(\pm 1)} \over k} = 0 \ .
\end{eqnarray}
Equations (\ref{veinst}) - (\ref{vdiv}) are not independent. Inserting Eq.(\ref{veinst}) into 
Eq.(\ref{veinst1}) we get Eq(\ref{vdiv}). The general solution of this system of equations is
\begin{eqnarray} \label{veqs}
B^{(\pm 1)} &=& \frac{k}{C_M}\frac{B^{(\pm 1)}_0}{a} \ , \\
H_T^{(\pm 1)} &=& \frac{B^{(\pm 1)}_0}{Ca_0^2}\left[\frac{k^2}{C_M}J(t)-\frac{2}{a_0}\coth(Ct)\right]+H^{(\pm 1)}_0 \ ,\nonumber
\end{eqnarray}
where $B_0^{(\pm 1)}(k)$, and $H_0^{(\pm 1)}(k)$ are arbitrary functions
depending on only spatial coordinates. The function $J(t)$ is also an elliptic integral 
of the time variable $t$, and can be expressed in terms of $I(t)$ as follows
\begin{eqnarray}
&&J(t)=-\frac{3}{2^{1/3}}I(t)-3{\rm sinh}^{-1/3}(Ct){\rm cosh}^{-1}(Ct)\ .
\end{eqnarray}\\
{\it Tensor modes}\\\\
For tensor modes we get a source-free gravitational wave propagation equation 
\begin{eqnarray} \label{teinst}
 a^2{{\ddot H}_T^{(\pm 2)}} + 3 { a \dot a}{{\dot H}_T^{(\pm 2)}} + k^2{H_T^{(\pm 2)}} = 0
\end{eqnarray}
as is the case in the absence of cosmological constant. The general solution of the wave 
equation in the presence of $\Lambda$ is given in Eq.(50) of \cite{PVCE}.

\section{The complete form of the metric}
Having in hand the solution for all modes of the perturbations, Eqs. (\ref{seqs}) and (\ref{veqs}), 
we can build up the general form of the metric (\ref{metric}).
The first order part of the metric takes the following form
\begin{eqnarray}\label{gab}
H_L\gamma_{\alpha\beta}&+&H_{T\alpha\beta}=H_T^{(+2)}Q_{\alpha\beta}^{(+2)}+
H_T^{(-2)}Q_{\alpha\beta}^{(-2)}\nonumber\\
&-&\frac{i}{2k}\left[\frac{B_0^{(+1)}}{Ca_0^2}\left(\frac{k^2J(t)}{C_M}
-\frac{2}{a_0}\coth(Ct)\right)+H_0^{(+1)}\right]\left(k_{\alpha}Q_{\beta}^{(+1)}
+k_{\beta}Q_{\alpha}^{(+1)}\right)\nonumber\\
&-&\frac{i}{2k}\left[\frac{B_0^{(-1)}}{Ca_0^2}\left(\frac{k^2J(t)}{C_M}
-\frac{2}{a_0}\coth(Ct)\right)+H_0^{(-1)}\right]\left(k_{\alpha}Q_{\beta}^{(-1)}
+k_{\beta}Q_{\alpha}^{(-1)}\right)\nonumber\\
&-&\left\{\left[\frac{a^3_0}{C_M}\coth(Ct)[K_1-K_2I(t)]-3V-\frac{3B_0}{2Ca_0^3}\coth(Ct)
+H_0\right]\frac{k_{\alpha}k_{\beta}}{k^2}\right.\nonumber\\
&+&\left.\left[\frac{B_0}{2Ca_0^3}\coth(Ct)-\frac{H_0}{3}\right]\gamma_{\alpha\beta}\right\}Q^{(0)} \ ,
\end{eqnarray}
and
\begin{eqnarray}\label{g0a}
g_{0\alpha}=-\frac{ka}{C_M}\left[B_0Q_{\alpha}^{(0)}+B_0^{(+1)}Q^{(+1)}_{\alpha}
+B_0^{(-1)}Q^{(-1)}_{\alpha}\right] \ .
\end{eqnarray}
To show the equivalence of the metric (\ref{gab}) and (\ref{g0a}) with the $C^{\infty}$ solution in \cite{PVCE},
we choose the space dependent integrational functions as follows,
\begin{eqnarray} \label{constants}
K_1 &=& -\frac{k^2C_M}{2a_0^3}A_k \ , \qquad\quad K_2=-\frac{3k^2C_M}{2^{7/3}a_0^5C^2}B_k \ ,  \nonumber\\
B_0 &=& -kC_M C_k^{(0)}=0 \ , \quad B^{(\pm 1)}_0 = -kC_MC_k^{(\pm 1)} \ ,\\ 
H_0 &=& 3V=\frac{3}{2}B_k\ , \quad\quad \ \ \ H^{(\pm 1)}_0 = 0 \ . \nonumber 
\end{eqnarray}
Here $A_k$, $B_k$, $C_k^{(0)}$ and $C_k^{(\pm 1)}$ are the components of the $k$th eigenmode of the 
spatial functions in Eqs. (52-54) of \cite{PVCE}. Inserting the values in Eq.(\ref{constants}) into 
Eqs. (\ref{gab}) and (\ref{g0a}) we find the agreement of the two solutions up to the remaining gauge 
freedom, which is not completely fixed in \cite{PVCE}.

\section{Concluding remarks}
The result of our work is relevant in the context of studying the CMBR fluctuations in the picture 
of first order approximation of a concordance model. We presented the complete form of the metric 
perturbations with cosmological constant in Bardeen's covariant linear formalism. We concluded that 
all the analytic solution can be obtained in agreement with the Sachs \& Wolfe approach. This result
can be useful in predicting the power spectrum of the fluctuations in the microwave background radiation.

We dedicate this paper to the memory of our supervisor Prof. Zolt\'an Perj\'es who took part actively in the
first period of the calculations. Regrettably we had to finish writing the manuscript without his help and
suggestions. 

\section{Acknowledgments}
This work was supported by OTKA grant no. TS044665.

\end{document}